\documentclass[preprint]{aastex}
\usepackage{epsf}

\def\kms{\ifmmode {\, \rm km \, s^{-1}}\else {$\, \rm km \, s^{-1}$}\fi }
\def\rsun{\ifmmode {\rm R_{\odot}}\else $\rm R_{\odot}$\fi}
\def\msun{\ifmmode {\rm M_{\odot}}\else $\rm M_{\odot}$\fi}

\def\linebreak{\hfil\break}

\def\\{\hfil\linebreak}
\def\
{\hfil\linebreak}

\def\msun{\ifmmode {\rm M_{\odot}}\else $\rm M_{\odot}$\fi}
\def\MSUN{\ifmmode {\rm M_{\odot}}\else $\rm M_{\odot}$\fi}
\def\msunyr{\ifmmode {\rm M_{\odot}\,yr^{-1}}\else $\rm 
M_{\odot}\,yr^{-1}$\fi}
\def\mdot{\ifmmode {\dot{M}}\else $\dot{M}$\fi}
\def\degree{\ifmmode {^\circ}\else {$^\circ$}\fi}
\def\mum{\ifmmode {\rm \mu {\rm m}}\else $\rm \mu {\rm m}$\fi}
\def\arcmin{\ifmmode ^{\prime}\else $^{\prime}$\fi}
\def\arcsec{\ifmmode ^{\prime \prime}\else $^{\prime \prime}$\fi}
\def\secpoint{\mbox{$''\mskip-7.6mu.\,$}}

\def\inch{\ifmmode ^{\prime \prime}\else $^{\prime \prime}$\fi}
\def\arcmin{\ifmmode ^{\prime}\else $^{\prime}$\fi}

\def\msun{\ifmmode {\rm M_{\odot}}\else $\rm M_{\odot}$\fi}
\def\lsun{\ifmmode {\rm L_{\odot}}\else $\rm L_{\odot}$\fi}
\def\mearth{\ifmmode {\rm M_{+\mskip-14.6muO\,}}\else $\rm 
M_{+\mskip-14.6muO\,}$\fi}
\def\mearth{\ifmmode {\rm M_{\earth}}\else $\rm M_{\earth}$\fi}
\newbox\grsign \setbox\grsign=\hbox{$>$} \newdimen\grdimen 
\grdimen=\ht\grsign
\newbox\simlessbox \newbox\simgreatbox
\setbox\simgreatbox=\hbox{\raise.5ex\hbox{$>$}\llap
     {\lower.5ex\hbox{$\sim$}}}\ht1=\grdimen\dp1=0pt
\setbox\simlessbox=\hbox{\raise.5ex\hbox{$<$}\llap
     {\lower.5ex\hbox{$\sim$}}}\ht2=\grdimen\dp2=0pt
\def\simgreat{\mathrel{\copy\simgreatbox}}
\def\simless{\mathrel{\copy\simlessbox}}
\begin{document}


\title{Extended Near-Infrared Emission from Candidate Protostars 
in the Taurus-Auriga Molecular Cloud}
\author{Shinae Park}
\affil{Physics Department, University of California,
366 LeConte Hall, Berkeley, CA 94720-7300}
\author{and}
\author{Scott J. Kenyon}
\affil{Smithsonian Astrophysical Observatory, 
60 Garden Street, Cambridge, MA 02138}
\received{17 December 2001}
\accepted{28 February 2002}
\author{to appear in}
\affil{The Astronomical Journal, June 2002}


\begin{abstract}
We describe near-IR imaging data for a sample of 23 class I sources
in the Taurus-Auriga dark clouds.  Combining our data with previous
photometry, we detect brightness variations of $\sim$ 0.1--0.5 mag in
many sources.  The near-IR morphologies are consistent with mm continuum
measurements.  Most ($\sim$ 60\%) of the sample are true protostars;
the rest may be objects in transition between class I and class II,
T Tauri stars with edge-on disks, or heavily reddened T Tauri stars.

\end{abstract}

\subjectheadings{infrared: stars --- stars: formation --- stars: 
pre-main-sequence --- ISM: reflection nebulae}

\section{INTRODUCTION}

Standard models for the formation of a single, low-mass star begin
with a slowly rotating cloud of gas and dust with a centrally 
concentrated density distribution 
\citep[e.g., Larson 1969, 1972; Stahler, Shu, \& Taam 1980, 1981;][]{Shu87,Shu93,saf97,cio00,whi01}.
When the inward gravitational pressure exceeds the outward thermal 
and magnetic pressures, the cloud collapses.  Material near the axis 
of rotation with low specific angular momentum contracts into a central 
star-like core.  Once the radius of the centrifugal barrier exceeds
the radius of the central stellar core, material away from the rotation 
axis with large specific angular momentum follows a curved infall trajectory,
falls onto the equatorial plane, and forms a circumstellar disk 
\citep{Cas81,Ter84,Lin90,Yor93,sta94}.  As the collapse proceeds, the 
protostellar core becomes more massive and more luminous; the
surrounding infalling envelope becomes less massive and less 
opaque at optical and infrared wavelengths \citep{Lada91}.  
Eventually, a wind from the star-disk system begins to clear 
gas and dust away from the rotational axis, resulting in 
the production of collimated jets and a bipolar molecular outflow
\citep{Ada93,Kon93,Kon00,Shu00}.  Once the infalling envelope disperses
or falls into the circumstellar disk, the central star-disk system  
becomes optically visible as a pre-main-sequence star.  As the 
material in the disk accretes onto the the central star, dissipates, 
or condenses into planets, the star moves onto the main sequence
\citep{Sta93,Str93,Cam95}.

This picture of star formation generally accounts for an apparent
evolutionary sequence in the observed properties of young stars in
nearby molecular clouds \citep{Lada84,Ada87,Lada87}.  \citet{Lada91}
describes three of the classes in this sequence.  Class I sources 
have spectral energy distributions (SEDs) which are broader than a 
single temperature blackbody function and rise from $\sim\!2~ \mu$m 
to $\sim\! 25\!-\!100~ \mu$m. Class I sources are star-disk systems 
still surrounded by massive infalling envelopes.  The envelope
absorbs optical and near-infrared photons from the central protostar
and re-emits this radiation at far-IR and mm wavelengths where the 
envelope is optically thin.  Class II sources also have broad SEDs, 
but the SED falls from $\sim\!2~ \mu$m to $\sim\! 25\!-\!100~ \mu$m.  
These sources contain a pre-main-sequence star surrounded by 
a circumstellar disk, which absorbs and reradiates emission 
from the central star.  Class III sources have roughly 
blackbody SEDs that peak at $\sim$ 1 $\mu$m; radiation from a surrounding 
disk or envelope contributes negligible emission at 10--100 $\mu$m 
(Lada 1991).  \citet{And93} added the class 0 sources to this 
sequence to include very young objects at the beginning of the 
accretion process.  These objects have narrow SEDs which peak at
$\sim$ 100--300 $\mu$m \citep[e.g.,][and references therein]{mot01}.  
In the current picture, the central star of a class 0 source has
accumulated $\lesssim$ 50\% of its final mass; in class I sources,
the central star has $\gtrsim$ 50\% of its final mass.

The Taurus-Auriga cloud is one of the best sites for testing models 
of isolated, low mass star formation 
\citep[see][and references therein]{kh95}.  This collection of 
filamentary clouds contains good samples of class I, II, and III
pre-main sequence stars, a class 0 source, and several examples
of interesting transition objects.  The clouds are close enough
\citep[d = 140 pc;][]{ken94a} to allow high signal-to-noise observations 
at all wavelengths with moderate to high spectral and spatial resolution.  
Although the clouds are poor examples of clustered star formation,
pre-main sequence stars in Taurus-Auriga provide interesting constraints 
on the formation of small close groups of pre-main sequence stars 
\citep{gom93,arm97,van98,sca99}.

Here, we consider a comparison between the millimeter (mm) and near-infrared
(near-IR) properties of class I sources in Taurus-Auriga.  \citet{mot01}
demonstrate that moderate spatial resolution mm continuum observations
divide these objects into unresolved, point-like sources with negligible 
infalling envelopes, extended sources with significant infalling envelopes,
and several transition objects with modest infalling envelopes. In this
division, the unresolved Taurus-Auriga sources might be edge-on disks
similar to those discovered in HH30 IRS and HK Tau \citep{bur96,sta98}.  
Because the derived lifetimes in each evolutionary phase depend on the 
relative numbers of objects in each phase \citep[e.g.][]{ken90,hog01}, 
the mm continuum data also provide better
limits on the class I lifetime in Taurus-Auriga.  Our goal is to test
whether near-IR images yield a similar division of sources into extended
and point-like objects.  This study extends previous imaging surveys, 
where near-IR imaging data constrains radiative transfer models for the 
class I sample \citep{ken93b,whi97}.

We describe the observations in \S2, comment on source variability
in \S3, and consider the image morphology in \S4.  The paper concludes
with a brief summary.

\section{OBSERVATIONS}

We acquired JHK images of candidate protostars in Taurus-Auriga with
Stelircam at the Fred L. Whipple Observatory 1.2-m telescope on
5--12 October 1998 and 18--20 November 1999. Stelircam is a dual-beam
IR imager equipped with two 256 $\times$ 256 InSb SBRC arrays. The
field-of-view for these images is $\sim$ 300\arcsec~$\times$ 300\arcsec.
We took nine 6 $\times$ 10 sec exposures of each field at HK and 
later at JK and offset the telescope by 45\arcsec--60\arcsec~after 
each integration.  Poor weather prevented a complete set of J band 
observations in photometric conditions; the 1998 HK observations were 
photometric.  We repeated observations of some targets to improve
the signal-to-noise of faint nebulosity.

We produce calibrated object frames in each filter using standard
techniques in the Interactive Data Language (IDL)\footnote{IDL
is distributed by Research Systems, Inc. of Boulder, Colorado.}.  
After correcting each image for non-linearity, 
we construct a nightly flat-field for each filter from the median 
of dark-subtracted image frames.  The sky frames for each object 
image are a median-filtered set of flattened frames scaled to have 
the same background as the object image.  We offset and coadd the
nine frames in each filter to make a final deep field image which
reveals fainter extended emission without saturating bright stars. 

Table 1 lists HK photometry for each source. We use the NOAO 
IRAF\footnote{IRAF is distributed by the National Optical 
Astronomy Observatory, which is operated by the Association of 
Universities for Research in Astronomy, Inc. under contract to the
National Science Foundation.}
routine PHOT to extract magnitudes in three apertures.  We analyze the 
nine HK frames of each source -- instead of the coadded frames --
to obtain error estimates for the photometry directly from the data.
Once we remove bad pixels, the typical standard deviation of nine
measurements for one source is $\pm$0.01--0.02 mag at K and
$\pm$0.02--0.03 mag at H. Our
nightly photometric calibration is based on 15--25 observations of
\citet{eli82} standard stars.  The uncertainty in the calibration
is $\pm$0.03 mag in each filter on each night. This uncertainty is 
comparable to the range in the zero-points and airmass corrections 
for the HK filters on each observing run.  Adding the calibration 
error in quadrature with the standard deviations of the source data
yields a typical photometric error of less than 0.05 mag in each filter.

To measure morphological information from the individual and coadded
frames, we use Sextractor \citep{ber96}.  Sextractor is a modern 
software package which automatically detects, measures, and classifies  
sources from digital images.  For each image,  we set up the package 
to measure the FWHM $f$, elongation $e$, central surface brightness
$\mu$, and the magnitude in 7\arcsec~and 14\arcsec~apertures for each 
candidate protostar.  Table 2 lists the results. 
The FWHM is the full width at half maximum of 
the best-fitting one dimensional gaussian to the intensity profile of
the source; the elongation is the ratio of the major axis to the minor 
axis of the best-fitting two dimensional gaussian to the intensity profile. 
The fits for both parameters use intensity levels 3$\sigma$ or more
above the local noise background.  We use the central surface brightness 
and aperture magnitudes to verify that fits for each source reach the same
limiting surface brightness and to check the PHOT photometry.  The
Sextractor photometry agrees with PHOT to $\pm$ 0.02 mag or better.
The limiting surface brightness in the coadded frames is $\sim$ 
19.5 mag arcsec$^{-2}$.

Figure 1 gives examples of contour maps derived from coadded K band images 
for Taurus-Auriga protostars
\citep[see also][]{Tam91,ken93b,whi97,luc97}.  The images have a variety 
of morphologies, ranging from apparent point sources (04108+2803 and 
04295+2251) to monopolar (04287+1801 and 04288+1802) and bipolar 
(04325+2402 and 04361+2547) nebulae with central point sources.
Two objects are faint patches with no obvious point source 
(04166+2706 and 04368+2557).  \citet{whi97} measure large
near-IR polarizations for sources with extended nebulae, confirming
models where dust grains within the nebulosity scatter radiation from the
central star into our line-of-sight \citep[see also][]{luc98a}.  
\citet{gom97} identify optical
and near-IR jets on [S~II] and H$_2$ images of many Taurus-Auriga
 protostars \citep[see also][]{luc98b}; 
\citet{ken98} note [S~II] emission on optical spectra of several other
sources.  Jet emission and bipolar outflows are characteristic of
embedded protostars \citep[e.g.,][]{rei91,mor92,Bon96,Tam96,rei01}.

The IRAS source 04166+2706 deserves special mention.  \citet{ken90} 
discovered a near-IR counterpart with raster scans on the single 
channel MMT photometer.  From near-IR imaging data, \citet{ken93b} 
identified a point source with very red colors near the IRAS position.  
This object is $\sim$ 2 mag brighter than the \citet{ken90} source 
and lies near the reddening band in the J--K,H--K color-color
diagram.  Our images recover the \citet{ken90} source as a patch of 
low surface brightness nebulosity similar to the nebula of 04368+2557
\citep[see][]{whi97}.  This object is 52\arcsec~N and 44\arcsec~W of the 
\citet{ken93b} point source, with J2000 coordinates of
RA = 04$^{\rm h}$19$^{\rm m}$43.2$^{\rm s}$ and 
Dec = +27\degree13\arcmin39\arcsec.  
The error in these coordinates is roughly $\pm$ 2\arcsec. 
We suggest that this object is the IRAS counterpart.

The triplet of embedded sources associated with IRAS sources 04181+2654 
and 04181+2655 also merit a short comment.  \citet[][1993b]{ken93a} 
associate the northern component with IRAS 04181+2655 and the southern 
pair with IRAS 04181+2654.  In this convention, the southern 
embedded source is 04181+2654A; the middle component of the three
is 04181+2654B.  Figure 1 labels these three sources.  For completeness,
the J2000 coordinates are
RA = 04$^{\rm h}$21$^{\rm m}$11.6$^{\rm s}$ and 
Dec = +27\degree01\arcmin10\arcsec~for 04181+2654A,
RA = 04$^{\rm h}$21$^{\rm m}$10.5$^{\rm s}$ and 
Dec = +27\degree01\arcmin38\arcsec~for 04181+2654B, and
RA = 04$^{\rm h}$21$^{\rm m}$08.1$^{\rm s}$ and 
Dec = +27\degree02\arcmin22\arcsec~for 04181+2655.

\section{Variability}

Most pre-main sequence stars vary at optical and infrared wavelengths
\citep[e.g.,][]{joy45,bou93,cho96,sta99}.  
Many systems have regular variations with amplitudes of 0.1--1.0 mag 
and periods of 2--10 days \citep[][2000]{eat95,her94}.  Other objects 
vary irregularly with similar amplitudes on similar or longer timescales.  
The periodic
variations yield information on stellar rotational periods and the 
sizes, temperatures, and lifetimes of dark and bright spots on the
stellar photosphere \citep{bou93,sta99}.  These data also provide 
tests of models for the inner disks and magnetospheres of low mass 
pre-main sequence stars \citep[e.g.,][]{mah98,woo98,woo00}.

To test whether any of the candidate protostars are variable, we 
compare with previous measurements obtained with single channel
photometers \citep{ken90} and near-IR imagers \citep{ken93b,ken94b,whi97}.  
For sources where we have only one or two of our own measurements,
we supplement the photometry with data from the compilation of
\citet{kh95}.
We derive the average brightness and the standard deviation for the 
candidate protostar and for another stellar source in the same field 
using 8\arcsec~and 13\arcsec--14\arcsec~apertures.  Although there 
are too few measurements for each source to construct a robust 
average, the standard deviation provides a reasonable first estimate 
of source variability.  

The observations in this study use different instruments; we use 
the comparison stars to verify that the photometry from each 
instrument is on the same magnitude scale.  We could not perform 
this test with single channel photometry; previous comparisons show 
that the differences between photometric systems are small compared 
to the source variations in our sample \citep{kh95}.  As a final test, 
we check whether the mean variation of a protostar from one observing 
run to the next is zero.  Using data from \citet{ken93b}, \citet{whi97}, 
and this paper, the average change in brightness of the complete 
sample is 0.11$\pm$0.37 from 1991 to 1993--94 and 0.05$\pm$0.31 
from 1993--94 to 1998--99.  

Our analysis indicates that candidate protostars in Taurus-Auriga are 
variable.  Histograms of standard deviations $\sigma$ in K and in H--K 
(Figure 2) demonstrate that candidate protostars have larger variations
than comparison stars in the same field.  The variation of a typical 
source has an amplitude of 0.2--0.3 mag at K.  These fluctuations are 
similar to variations observed in optically visible T Tauri stars
\citep[e.g.,][]{kh95,cho96,sta99}.  Many sources
have variable colors.  Roughly half of the sample varied by 0.1--0.3
mag in H--K.  Although large color variations are observed in some 
T Tauri stars, the variations are less than 0.1--0.2 mag in more than 
70\% of pre-main sequence stars in Taurus-Auriga 
\citep[][see also Carpenter, Hillenbrand, \& Skrutskie 2001]{kh95}.

Without repeat observations on short timescales, we are unable to 
analyze the variability of Taurus-Auriga protostars in detail.  
The large brightness and color changes are consistent with the 
behavior expected from bright spots, dark spots, and obscuration models
\citep[see also][and references therein]{car01}.  There is no 
correlation between brightness and color changes; some sources 
become bluer as they brighten while others become redder.
Most models predict strong correlations between color and brightness
evolution, but our observations are too few in number to discriminate 
between possible models.

The variations of several objects deserve special mention.

\noindent
04018+2803B -- This source is the more embedded of a pair of {\it IRAS}
sources separated by 17\secpoint5 on the sky.  The large variation
is based on a single observation in October 1998.

\noindent
04181+2655 -- The brightest of a triplet of {\it IRAS} sources, this
object has varied consistently from one observing run to the next.
The color variation of this source is among the largest in our sample.

\noindent 
04239+2436 -- This source has a large nebulosity at J which is nearly
invisible at K.  It has one of the largest amplitudes in our sample, 
with a variation of more than 0.7 mag at K during the past decade.

\noindent
Six other sources -- 04016+2610, 04181+2654A, 04287+1801 (L1551 IRS5), 
04361+2547, 04365+2535, and 04381+2540 -- vary by 20\% or more in brightness 
and in color.  All other sources change by less than 0.1--0.2 mag 
(Table 1; columns 10--11).  The historical variation of L1551 IRS5 is 
difficult to quantify due to large differences in observing techniques, 
but published data are consistent with the variation in our data.

\section{Morphology}

Image morphology provides a good constraint on models for embedded 
protostars.  \citet{whi97} demonstrate that the size of synthetic 
near-IR images correlates with the centrifugal radius $R_c$ and the 
infall rate $\dot{M}$ in the \citet{Ter84} infalling envelope model. 
The elongation correlates with the inclination of the source and the 
size of the outflow cavity \citep[see also][]{whi93}.  Both parameters 
vary with the grain albedo. Grains with large albedo produce larger 
and more elongated near-IR images of protostars.  \citet{whi97} 
emphasize that the size and elongation provide better constraints 
on models when combined with polarization measurements.

The image morphology also provides a good comparison with mm continuum
observations.  \citet{mot01} describe moderate spatial resolution mm-wave 
observations of the Taurus-Auriga sample.  They divide the sources into 
two groups based on the ratio of the flux $F$ in 11\arcsec~and 
60\arcsec~beams.  Extended sources have $F_{11}/F_{60} \lesssim$ 0.5--0.7; 
point sources have $F_{11}/F_{60} \approx$ 1. 
Roughly half of the Taurus-Auriga sample is extended by this criterion.

To compare the near-IR morphology with mm continuum morphology, we use two
measures to divide objects into extended sources and point-like sources.
We follow \citet{mot01} and define extended sources as objects where
the flux in a 26\arcsec~aperture exceeds the fluxes in 13\arcsec~and
8\arcsec~apertures (Table 1).  Larger apertures yield poor quality 
photometry on our images. In a sample of $\sim$ 2000 field stars with 
moderate signal-to-noise photometry, the mean aperture corrections for 
our data are 
$m_{8}$--$m_{13}$ = 0.09$\pm$0.02 and $m_{13}$--$m_{26}$ = 0.11$\pm$0.02.
The aperture corrections are independent of the filter.  With this result, 
extended sources have $m_{8}$--$m_{13} > $ 0.15 and $m_{13}$--$m_{26} > $ 0.17.  
The size (FWHM) $f$ and elongation $e$ also provide good measures of source 
extension.  In a sample of $\sim$ 3000 moderate signal-to-noise detections 
from our K-band images, the median size is $f_{med}$ = 
1\secpoint75$\pm$0\secpoint30 where the error bar refers to the 
inter-quartile range (Figure 3).  The median elongation in the K-band images 
is $e_{med}$ = 1.10$\pm$0.04.  Point sources are well-defined in our images.  
We thus define extended sources to have $f_K >$ 2\secpoint5 and $e_K >$ 1.2.

Our analysis indicates a good correlation between various measures
for extended emission in the Taurus-Auriga class I sample (Figure 4).
For sources with $f_K \le$ 5\arcsec, there is a one-to-one relation 
between $f_K$ and $K_{8}$--$K_{13}$ (Figure 4; lower left panel).
Although this relation is weaker at larger sizes, the Spearman rank
order test \citep{Press88} still indicates a significant correlation with 
a probability, $p_{s} \approx$ 3 $\times 10^{-4}$, of being drawn
from a random sample.  The elongation and size derived from Sextractor
show a stronger correlation (Figure 4, lower right panel), with 
$p_{s} \approx$ 5 $\times 10^{-5}$.

The good correlation between size and elongation confirms predictions 
of radiative transfer models for protostars with infalling envelopes.
In the \citet{Ter84} infall models with a bipolar outflow cavity, the size
and the elongation of near-IR images are related to infall rate and the 
geometry of the outflow cavity.  Protostars with denser envelopes, 
larger infall rates, and curved or dust-filled outflow cavities are 
larger and more elongated than protostars with less dense envelopes,
smaller infall rates, or streamline cavities \citep{ken93b,whi97}.  Our data 
confirm the general correlation predicted by the models.  Further comparisons 
between the data and the models will yield better tests of the models.

The K-band size and elongation are well-correlated with the mm continuum 
measurements.  \citet{mot01} define the concentration $c$ as the ratio of 
the fluxes in 11\arcsec~and 60\arcsec~apertures expressed as a percentage.
Highly concentrated
sources have $c$ = 100; very extended sources have $c < $ 30.
Our near-IR measures $f_K$ and $e_K$ are inversely correlated with
$c$ (Figure 4, middle panels).  The relation between $c$ and $e_K$ 
is significant at the 3$\sigma$ level, with
$p_{s} \approx$ 4 $\times 10^{-3}$; the size is more weakly correlated
with $c$ ($p_{s} \approx$ $10^{-2}$).  \citet{mot01} use the concentration 
and the absolute fluxes to measure the envelope mass, $M_{env}$.  
Our K-band measures are more strongly correlated with $M_{env}$ 
than with $c$ (Figure 4, upper panels); 
$p_{s} \approx$ 2 $\times 10^{-3}$ is the Spearman rank probability
for the correlation between $f_K$ and $M_{env}$ and
$p_{s} \approx$ $10^{-3}$ is the probability for the correlation 
between $e_K$ and $M_{env}$.

The near-IR measurements at J and H also correlate with the mm continuum 
data.  Because the short wavelength images have lower signal-to-noise, the 
correlations at J and H are 10\% to 20\% weaker than at K.  These 
results exclude the deeply embedded source 04368+2557 (L1527 IRS)
from the analysis. Including this source with a nominal size of 
$f_K \approx$ 25\arcsec~and an elongation of $e_K \approx$ 2 yields
smaller Spearman rank probabilities by a factor of 2--3.

These correlations provide additional confirmation of infall models for
protostars. In the standard infall picture \citep{Ter84,Ada87}, an 
optically thick envelope absorbs radiation from the central object and 
re-radiates this energy at mm wavelengths.  Protostars with massive 
infalling envelopes are therefore more extended at mm continuum 
wavelengths than class II or class III sources with little or no 
infalling envelope.  Once the central object generates a bipolar 
outflow, optical and near-IR radiation from the central object can 
escape through the optically thin cavity \citep[e.g.][]{whi93,ken93b,whi97}.
Some of this radiation scatters off material in the cavity and in the 
outer envelope, producing an extended, elongated, and highly polarized
optical or near-IR image \citep{whi97}. As material from the envelope 
continues to fall onto the disk, the envelope opacity declines and more 
optical and IR radiation from the central star can reach the observer 
directly. This process reduces the amount of mm continuum emission and 
scattered optical 
and near-IR emission.  These objects thus appear more concentrated and 
more symmetrical at all wavelengths than more highly embedded objects.
Our data confirm this general trend: class I sources that are extended
at mm continuum wavelengths are extended and more elongated on near-IR 
images than their point-like counterparts.

These correlations may also offer a test of infalling envelope models.
To the best of our knowledge, there are few quantitative predictions of
the relationship between the optical or near-IR properties and the
mm continuum properties of protostars.  Published radiative transfer
models have successfully modeled independently the spectral energy
distributions, density distributions, and near-IR images of individual
protostars.  However, we are not aware of models that predict the 
observed correlations between the concentration or envelope mass derived 
from mm data and the size or elongation derived from optical or near-IR 
data.  Our results suggest that observations can now test these predictions.

The near-IR data divide Taurus-Auriga class I sources into nearly the 
same two groups as the mm continuum data. We base this conclusion 
on Table 3, which compares our results with those of \citet{mot01}.
Nine sources are extended at mm and near-IR wavelengths;
eight sources are not extended at these wavelengths. Two sources --
04016+2610 and 04302+2247 -- have large near-IR nebulae with little
evidence for extended millimeter continuum emission from a massive
envelope\footnote{04016+2610 is resolved at 2 cm and 3.5 cm 
\citep{luc00}; this emission probably traces radiation from an 
ionized jet or inner disk as in L1551 IRS5 \citep{rod86}.} 
\citep{hog01,mot01}.  Three sources -- 04181+2654A, 04181+2654B, 
04365+2535, and 04381+2540 -- are extended at mm continuum wavelengths
but are not extended at near-IR wavelengths based on our measurements
of size and elongation.  Three of these -- 04181+2654A, 04365+2535, 
and 04381+2540 --
show some evidence for an extended nebula at K from $K_{8}$--$K_{13}$ 
and a visual inspection of the coadded images (see Figure 1).  Higher 
signal-to-noise K-band images 
would provide definitive tests for extended emission in these sources.

Our data confirm the \citet{mot01} conclusion that most ($\sim$ 60\%) 
of the Taurus-Auriga class I sources are embedded protostars with massive 
infalling envelopes \citep[see also][and references therein]{hog98,hog00,cha00}.  
Because these objects are definitely
extended in the mm continuum and probably extended at K, we include 
04181+2654A, 04181+2654B, and 04381+2540 in this group (see Table 2).
The envelope masses inferred from the millimeter data range from $\sim$
0.09 $M_{\odot}$ for 04239+2436 to $\sim$ 0.9 $M_{\odot}$ for L1551 IRS5.
The density distributions for these sources, $\rho(r) \propto r^{-n}$
with $n$ = 1.5--2, agree with the predictions of infall models on 
scales of $\sim$ 10,000 AU \citep{Ter84,cha98,hog98,hog00,mot01}.  The SEDs 
and the near-IR morphologies and 
polarizations of these sources also agree with model predictions 
\citep[Kenyon et al. 1993a,b;][]{whi97,mot01}.  All of these sources
have molecular outflows \citep[e.g.,][]{Bon96} and 66\% have emission 
from optical or near-IR jets \citep[e.g.,][]{gom97}. 

At least two Taurus-Auriga class I sources without extended mm emission 
might be transition objects between class I protostars and class II 
T Tauri stars.  In addition to their large near-IR nebulae, 04016+2610 
and 04302+2247 have molecular outflows, optical or near-IR emission 
from a jet, and large near-IR polarization as in true protostars
\citep[see also][]{Hey90,Tam91,ken93b,gom97,whi97}.  However, both
sources fail to meet the ``true protostar'' criterion of \citet{mot01}, 
$M_{env}/M_{\star} \simgreat$ 0.1.  To estimate these ratios in 
04016+2610 and 04302+2247, we assume the central star in each source 
lies on the birthline of \citet{Pal99}.  The observed bolometric 
luminosities then yield approximate stellar masses of $M_{\star} \sim$ 
0.6 $M_{\odot}$ for 04016+2610 and $M_{\star} \simless$ 0.1 $M_{\odot}$ 
for 04302+2247.  \citet{mot01} estimate envelope masses 
$M_{env} \simless$ 0.04  $M_{\odot}$ for 04016+2610 and
$M_{env} \simless$ 0.01  $M_{\odot}$ for 04302+2247.
Thus, both sources have $M_{env}/M_{\star} \simless$ 
0.1 and are not true protostars.

Based on mm aperture synthesis observations, \citet{hog01} previously 
proposed that 04016+2610 is a transition object between true protostars
and class II sources. The mm data are consistent with a Keplerian disk 
with a radius of $\sim$ 2000 AU surrounding a 0.65 $M_{\odot}$ central
star \citep[see also][]{sai01}.  Our mass estimate favors this 
interpretation over the proposal of
sub-Keplerian motion around a more massive star.  If 04016+2610 is a
transition object, several common near-IR properties suggest that 
04302+2247 might also be a transition object\footnote{\citet{pad99} 
detect an edge-on disk on HST near-IR images of 04302+2247, but our 
ground-based near-IR images of 04302+2247 show more near-IR nebulosity 
than class II sources with edge-on disks such as HH30 IRS and HK Tau.}.
Aperture synthesis observations at mm wavelengths would test this 
proposal.

Other Taurus-Auriga class I sources are peculiar in having class I SEDs 
but little or no extended near-IR or mm continuum emission.  The properties 
of this group differ significantly from the true protostars.  Most do 
not have molecular outflows, optical/near-IR jet emission, or large
near-IR polarization.  As a group, these objects thus appear older,
with smaller extended envelopes and less accretion activity than the 
protostars \citep[see also][]{rei01}.  There are several exceptions 
to this statement.  Four 
sources -- 04158+2805, 04260+2642, 04264+2433, and 04489+3042 --
have [S~II] emission; 04264+2433 has large near-IR polarization; 
near-IR spectra of 04295+2251 and 04489+3042 more closely resemble 
spectra of class I sources than T Tauri stars \cite{gre96}. This 
diversity of properties suggests that the peculiar class I sources 
do not belong to a single class of objects.  

Table 2 lists our best estimates for the nature of these peculiar class
I sources.  We consider 04016+2610, 04260+2612, 04264+2433, 04302+2247 
as transition objects between class I and class II.  Extended, highly
polarized near-IR nebulae demonstrate that 04016+2610 and 04302+2247 are
not class II sources; 04181+2655, 04260+2612 and 04264+2433 have similar 
spectral energy distributions but no obvious near-IR nebulosity.  
The class I sources 
04108+2803B, 04295+2251, and 04489+3042 have larger optical 
to far-IR flux ratios than 04016+2610 and other `transition candidates.'  
These sources could be transition objects viewed along the rotational 
axis of the disk. The strong [S II] emission in 04295+2251 recalls 
similar emission in the edge-on disk systems HK Tau and HH30 IRS.
Finally, 04108+2803A and 04158+2805 are more similar to class II sources 
and are probably heavily reddened T Tauri stars.  

\section{SUMMARY}

Our near-IR observations provide a first detection of near-IR variability 
in a sample of class I protostars.  The variations are similar in amplitude 
to those observed in T Tauri stars \citep{kh95,car01}.  The poor time
coverage of the imaging data for this sample does not provide a useful 
measure of the timescale for these variations.  If the variations are
similar to those in T Tauri stars, observations on successive nights
over weeks or months can measure the timescales.  As an example, 
\citet{hil01} derive an 8 day period for the variation in the
Becklin-Neugebauer object, an embedded protostar in Orion 
\citep[see also][]{woo00}.

The near-IR morphologies of Taurus-Auriga class I sources are consistent
with the mm continuum measurements of \citet{mot01}.  Most ($\sim$ 60\%)
of these
objects are true protostars, with massive envelopes surrounding a
young star with $\sim$ 50\% of its final mass.  
Some ($\sim$ 25\%) of the class I sources are probably transition 
objects between true protostars and optically visible T Tauri stars; 
one or two of these might have edge-on disks similar to those observed 
in HH30 IRS and HK Tau.  Two other class I sources are probably heavily 
reddened T Tauri stars;.

Making accurate classifications for the ``peculiar'' class I sources 
requires additional data.  Radio observations of some class I sources 
using higher spatial resolution might reveal compact nebulae or large 
disks as in 04016+2610 \citep{hog01}.  Deeper near-IR images may also 
reveal extended nebulosity.  We have marginal detections of faint 
nebulosity on coadded near-IR images of 04181+2655 and 04264+2433; 
higher signal-to-noise observations might provide a better indication 
of the extended nature of these and other point-like class I sources.  
Because molecular outflows and optical/near-IR jets are more common 
among class I sources and transition objects than among T Tauri stars 
with class II SEDs \citep{gom97,ken98,rei01},  deeper optical and 
near-IR surveys for jet emission and more extensive outflow surveys 
can also help to classify these objects.  These new data would provide 
a better understanding of the evolution from class I to class II among 
low mass protostars.  

Finally, our results offer a new test of infall models for class I protostars.
In the standard picture, a massive envelope produces mm and far-IR continuum 
emission; near-IR and optical radiation from the central object escape 
through the optically thin outflow cavity.  As the infalling envelope becomes 
less massive and less opaque, the angular 
size of the outflow cavity should shrink.  Because optical and near-IR 
radiation can escape the envelope more easily, protostars should become less 
elongated as the optical depth through the envelope decreases.  Our near-IR 
morphological data confirm this expectation: $f_K$ and $e_K$ of class I 
sources correlate with $M_{env}$.  The standard model also predicts a general
anti-correlation between the near-IR measurements $f_K$ and $e_K$ and the 
mm measurement $c$ of \citet{mot01}.  As the envelope mass declines, the
unresolved central star and the disk produce more of the mm continuum flux;
sources thus become more concentrated at mm wavelengths as they become
less elongated and less extended at optical and near-IR wavelengths. To
the best of our knowledge, however, theoretical radiative transfer models 
of protostars do not make specific predictions for the variation of $f_K$ 
or $e_K$ as a function of $c$ or $M_{env}$.  Our results demonstrate that 
near-IR and mm continuum data can test these predictions.

\vskip 6ex

We thank C. Lada for advice and comments on this project.  M. Geller
and an anonymous referee made many useful comments on the manuscript.
W. Wyatt and J. Grimes provided technical assistance. 
S. P.  thanks the National Science Foundation for their support of 
the REU program at the Smithsonian Astrophysical Observatory where 
most of this work was carried out. S. K. thanks Dr. S. Starrfield and
Arizona State University for their hospitality during the completion 
of this project.

\vfill
\eject
%
%
\begin{deluxetable}{ lccrcrcrccc}
\tabletypesize{\small}
\tablecaption{Photometry of Taurus-Auriga candidate protostars}
\label{tbl-1}
\tablewidth{0pt}
\tablehead{ & & & \multicolumn{2}{c}{$8 \, \arcsec$ aperture} & \multicolumn{2}{c}{$13 \, \arcsec$ aperture} & \multicolumn{2}{c}{$26 \, \arcsec$ aperture} \\
\colhead{IRAS Name}  & \colhead{Other ID} & \colhead{JD} & \colhead{K} & \colhead{H--K} & \colhead{K} & \colhead{H--K} & \colhead{K} & \colhead{H--K} &
\colhead{$\sigma_K$} & \colhead{$\sigma_{H-K}$}}
\startdata
04016+2610 & L1489 IRS & 51093.9 &  9.45 & 2.14 &  9.20 & 1.92 &  9.10 & 1.89 & 0.21 & 0.21 \\
04108+2803A& L1495 IRS & 51092.9 & 10.44 & 1.11 & 10.32 & 1.14 & 10.33 & 1.14 & 0.04 & 0.07 \\
04108+2803B& L1495 IRS & 51092.9 & 10.70 & 2.19 & 10.55 & 2.25 & 10.52 & 2.31 & 0.52 & 0.18 \\
04158+2805 & \nodata & 51092.9 & 11.03 & 1.04 & 10.90 & 1.02 & 10.86 & 1.02 & 0.16 & 0.06 \\
04166+2706 & \nodata & 51093.9 & 15.05 & \nodata & 14.27 & \nodata & 13.65 & \nodata & \nodata & \nodata \\
04169+2702 & \nodata & 51093.9 & 11.40 & 2.45 & 11.17 & 2.24 & 11.01 & 1.95 & 0.11 & 0.13 \\
04181+2655 & \nodata & 51092.9 &  9.89 & 1.54 &  9.76 & 1.50 &  9.72 & 1.49 & 0.45 & 0.31 \\
04181+2654B& \nodata & 51092.9 & 11.24 & 2.71 & 11.06 & 2.76 & 10.98 & 2.70 & 0.19 & 0.09 \\
04181+2654A& \nodata & 51092.9 & 10.52 & 2.23 & 10.31 & 2.25 & 10.24 & 2.20 & 0.24 & 0.09 \\
04239+2436 & \nodata & 51093.9 & 10.33 & 2.47 & 10.17 & 2.35 & 10.12 & 2.20 & 0.31 & 0.06 \\
04248+2612 & HH31 IRS2 & 51093.9 & 10.60 & 0.70 & 10.17 & 0.65 &  9.91 & 0.69 & 0.13 & 0.05 \\
04260+2642 & \nodata & 51093.9 & 11.53 & 1.14 & 11.45 & 1.15 & 11.42 & 1.17 & 0.27 & 0.13 \\
04264+2433 & Elias 6 & 51092.9 & 11.23 & 0.82 & 11.07 & 0.76 & 10.98 & 0.79 & 0.18 & 0.12 \\
04287+1801 & L1551 IRS5 & 51092.9 &  9.52 & 1.47 &  9.33 & 1.39 &  9.20 & 1.34 & 0.21 & 0.35 \\
04288+1802 & L1551 NE & 51092.9 & 11.22 & 2.06 & 10.97 & 1.98 & 10.79 & 1.97 & 0.08 & 0.06 \\
04295+2251 & L1536 IRS & 51092.9 &  9.90 & 1.71 &  9.77 & 1.69 &  9.68 & 1.74 & 0.15 & 0.05 \\
04302+2247 & \nodata & 51093.0 & 11.42 & 1.12 & 11.13 & 1.11 & 10.99 & 1.10 & 0.17 & 0.08 \\
04325+2402 & L1535 IRS & 51092.9 & 11.41 & 1.99 & 10.94 & 1.82 & 10.54 & 1.60 & 0.16 & 0.10 \\
04361+2547 & TMR1  & 51092.9 & 10.48 & 2.25 & 10.29 & 2.17 & 10.14 & 2.10 & 0.20 & 0.05 \\
04365+2535 & TMC1A & 51093.0 & 10.76 & 2.87 & 10.51 & 2.74 & 10.33 & 2.28 & 0.21 & 0.15 \\
04381+2540 & TMC1 & 51092.9 & 11.58 & 2.27 & 11.38 & 2.03 & 11.26 & 1.74 & 0.22 & 0.07 \\
04489+3042 & \nodata & 51094.9 & 10.37 & 1.56 & 10.28 & 1.52 & 10.25 & 1.46 & 0.15 & 0.13 \\
\enddata
\end{deluxetable}

\begin{deluxetable}{lccccccl}
\tablecaption{Morphological data for Taurus-Auriga Class I Sources$^a$} 
\label{tbl-2}
\tablewidth{0pt}
\tablehead{
\colhead{Source Name} & \colhead{$f_J$ (arcsec)} & \colhead{$e_J$} & 
\colhead{$f_H$ (arcsec)} & \colhead{$e_H$} & \colhead{$f_K$ (arcsec)} & \colhead{$e_K$} & \colhead{Comment$^{\rm b}$} }
\startdata                    
04016+2610  & \ 5.23    & 1.15 & 11.76 & 1.19  & \ 6.04  & 1.27 & transition \\
04108+2803A & \ 1.44    & 1.09 & \ 1.89  & 1.05  & \ 2.00  & 1.06 & reddened TTS \\
04108+2803B & \ 1.86    & 1.17 & \ 1.84  & 1.06  & \ 1.99  & 1.07 & transition? \\
04158+2805  & \ 1.93    & 1.21 & \ 1.90  & 1.10  & \ 2.10  & 1.10 & reddened TTS \\ 
04166+2706  & \nodata & \nodata  & \nodata & \nodata & \ 6.42  & 1.55  & protostar \\   
04169+2702  & \nodata & \nodata  & 14.40 & 2.43  & 13.35 & 1.98  & protostar \\    
04181+2655  & \nodata & \nodata  & \ 2.05  & 1.07  & \ 2.18  & 1.05  & transition? \\
04181+2654B & \nodata & \nodata  & \ 1.96  & 1.08  & \ 2.25  & 1.09  & protostar \\
04181+2654A & \nodata & \nodata  & \ 2.13  & 1.08  & \ 2.10  & 1.06  & protostar \\
04239+2436  & \ 4.94    & 1.36 & \ 2.27  & 1.06  & \ 3.19  & 1.19 & protostar \\
04248+2612  & 15.30   & 2.80 & 11.20 & 2.45  & 11.68 & 2.43 & protostar \\
04260+2642  & \nodata & \nodata  & \ 1.25  & 1.09  & \ 1.13  & 1.07 & transition? \\
04264+2433  & \ 1.52    & 1.07 & \ 1.77  & 1.05  & \ 1.95  & 1.04 & transition? \\
04287+1801  & \ 2.45    & 1.16 & 16.59 & 1.89  & 12.39 & 1.76 & protostar \\
04288+1802  & \ 4.17    & 1.48 & \ 2.77  & 1.18  & \ 3.63  & 1.34 & protostar \\
04295+2251  & \ 1.87    & 1.11 & \ 2.14  & 1.08  & \ 2.02  & 1.05 & edge-on disk? \\
04302+2247  & \ 4.26    & 1.48 & \ 3.35  & 1.27  & \ 3.63  & 1.29 & transition \\
04325+2402  & \nodata & \nodata & \ 3.62  & 1.76  & \ 9.75  & 1.69 & protostar \\
04361+2547  & \ 2.01    & 1.25 & \ 2.18  & 1.35  & \ 2.77  & 1.50 & protostar \\
04365+2535  & \ 1.57    & 1.09 & \ 2.28  & 1.18  & \ 2.44  & 1.32 & protostar \\
04368+2537  & \nodata & \nodata & \nodata & \nodata & $>$25.00 & $>$2.00 & protostar \\
04381+2540  & \nodata  & \nodata & \ 2.07  & 1.08  & \ 2.05 &  1.06 & protostar \\
04489+3042  & \nodata  & \nodata & \ 1.70  & 1.07  & \ 1.54  & 1.05 & transition? \\
\enddata
\tablenotetext{a}{The columns list the FWHM $f$ and elongation $e$ of gaussian fits to the intensity profile of each source at each wavelength. Based on multiple measurements of each source, the typical error in $f$ is $\pm$0\secpoint2; the typical error in $e$ is $\pm$0.1.}
\tablenotetext{b}{The `Comment' column lists our best estimate of the evolutionary class of each object, true protostars (protostar), objects in transition between class I and class II (transition), edge-on disks, or reddened T Tauri stars.}
\end{deluxetable}

\begin{deluxetable}{cccc} 
\tablecaption{Comparison of infrared and millimeter results}
\label{tbl-3}
\tablewidth{0pt}
\tablehead{
\colhead{extended} & \colhead{extended} & \colhead{extended in mm but not}  & \colhead{extended in near-IR but not} \\ \colhead{in both} & \colhead{in neither} &
	\colhead{in near-IR continuum} & \colhead{in mm continuum} 
}
\startdata    
\nodata & \nodata & \nodata   & 04016+2610 \\
\nodata & 04108+2803A & \nodata   & \nodata      \\
\nodata & 04108+2803B & \nodata   & \nodata      \\
\nodata & 04158+2805 & \nodata   & \nodata      \\
04166+2706 & \nodata & \nodata & \nodata      \\
04169+2702 & \nodata & \nodata     & \nodata      \\
 \nodata   & 04181+2655 & \nodata  & \nodata      \\
 \nodata   & \nodata & 04181+2654B & \nodata      \\
 \nodata   & \nodata & 04181+2654A & \nodata      \\
04239+2436 & \nodata   & \nodata   & \nodata      \\
04248+2612 & \nodata   & \nodata   & \nodata      \\
\nodata   & 04260+2642 & \nodata   & \nodata      \\
\nodata   & 04264+2433 & \nodata   & \nodata      \\
04287+1801 & \nodata   & \nodata   & \nodata      \\
04288+1802 & \nodata   & \nodata   & \nodata      \\
\nodata    & 04295+2251 & \nodata  & \nodata      \\
\nodata    & \nodata   & \nodata   & 04302+2247   \\
04325+2402 & \nodata   & \nodata   & \nodata     \\
04361+2547 & \nodata   & \nodata   & \nodata      \\
\nodata    & \nodata   & 04365+2535 & \nodata      \\
04368+2557 & \nodata   & \nodata   & \nodata      \\
\nodata   & \nodata    & 04381+2540 &\nodata      \\
\nodata   & 04489+3042 & \nodata   &  \nodata     \\
\enddata
\end{deluxetable}

\clearpage


\epsfxsize=8.0in
\hskip 2ex
\epsffile{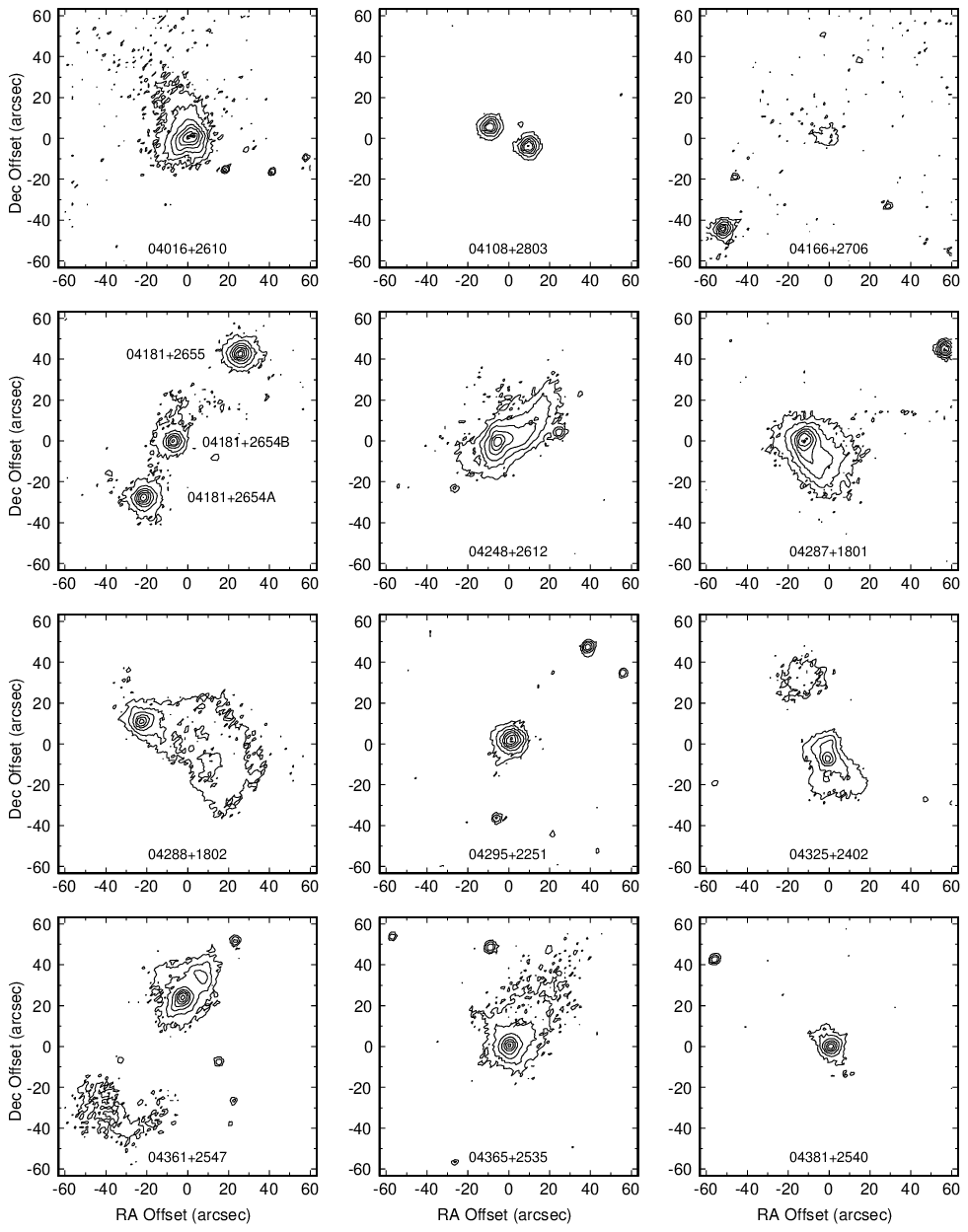}
\figcaption[SPark.fig1.eps]
{K-band contour maps for Taurus-Auriga class I sources. In each panel,
north is up and east is to the left. The lowest contour is $\sim$ 
3$\sigma$ above the background noise level; the contour spacing is 1 mag. }

\epsfxsize=7.0in
\hskip -10ex
\epsffile{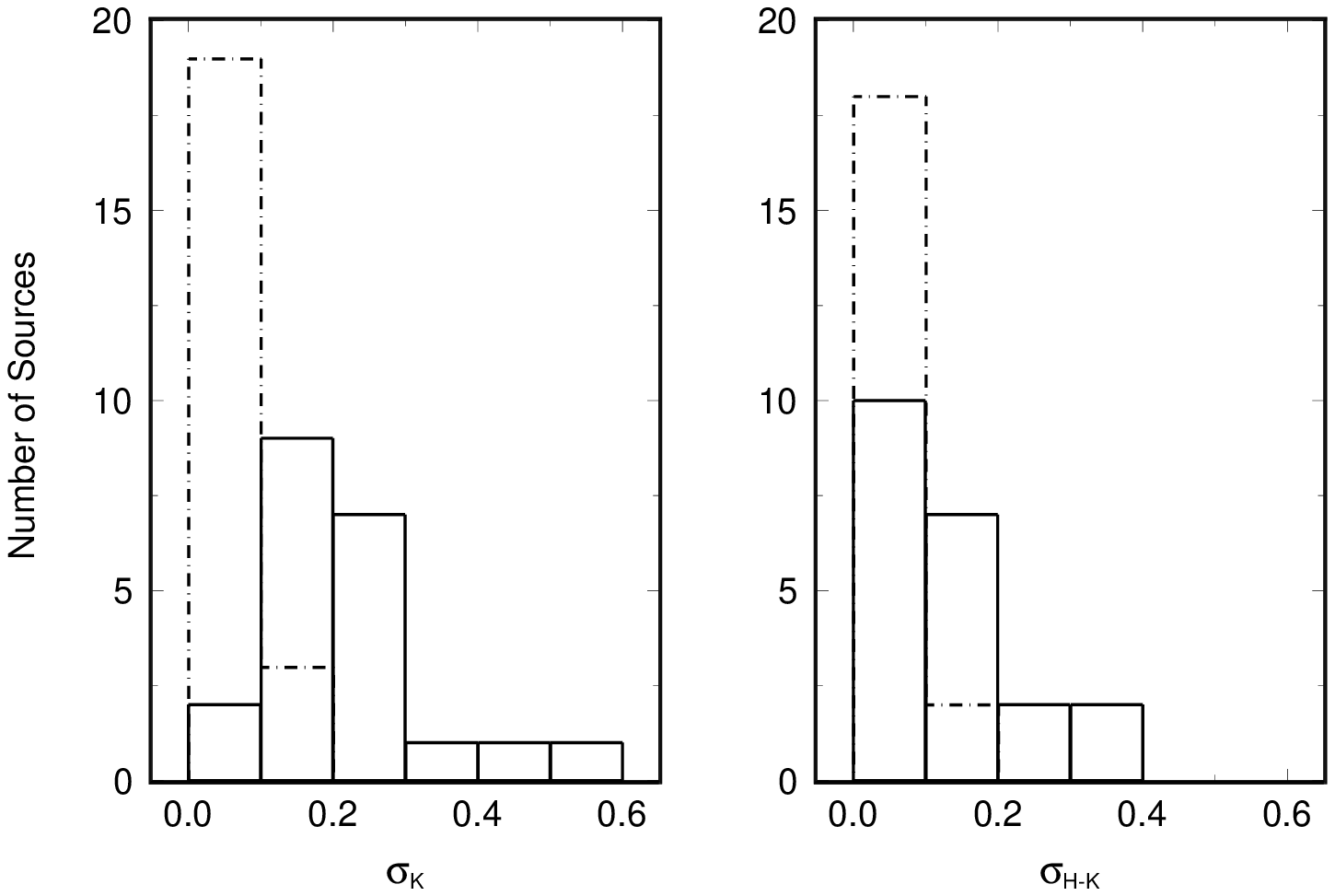}
\figcaption[SPark.fig2.eps]
{Variability of K and H--K in pre-main sequence stars.
The left panel shows the distribution of standard deviations 
$\sigma_K$ about the mean K magnitude for candidate protostars 
with 2 or more K measurements (solid line) and for comparison
stars in the same field (dot-dashed line).
The right panel shows the distribution $\sigma_{H-K}$ for H--K 
of candidate protostars (solid line) and of comparison stars 
(dot-dashed line).}

\epsfxsize=7.0in
\hskip -10ex
\epsffile{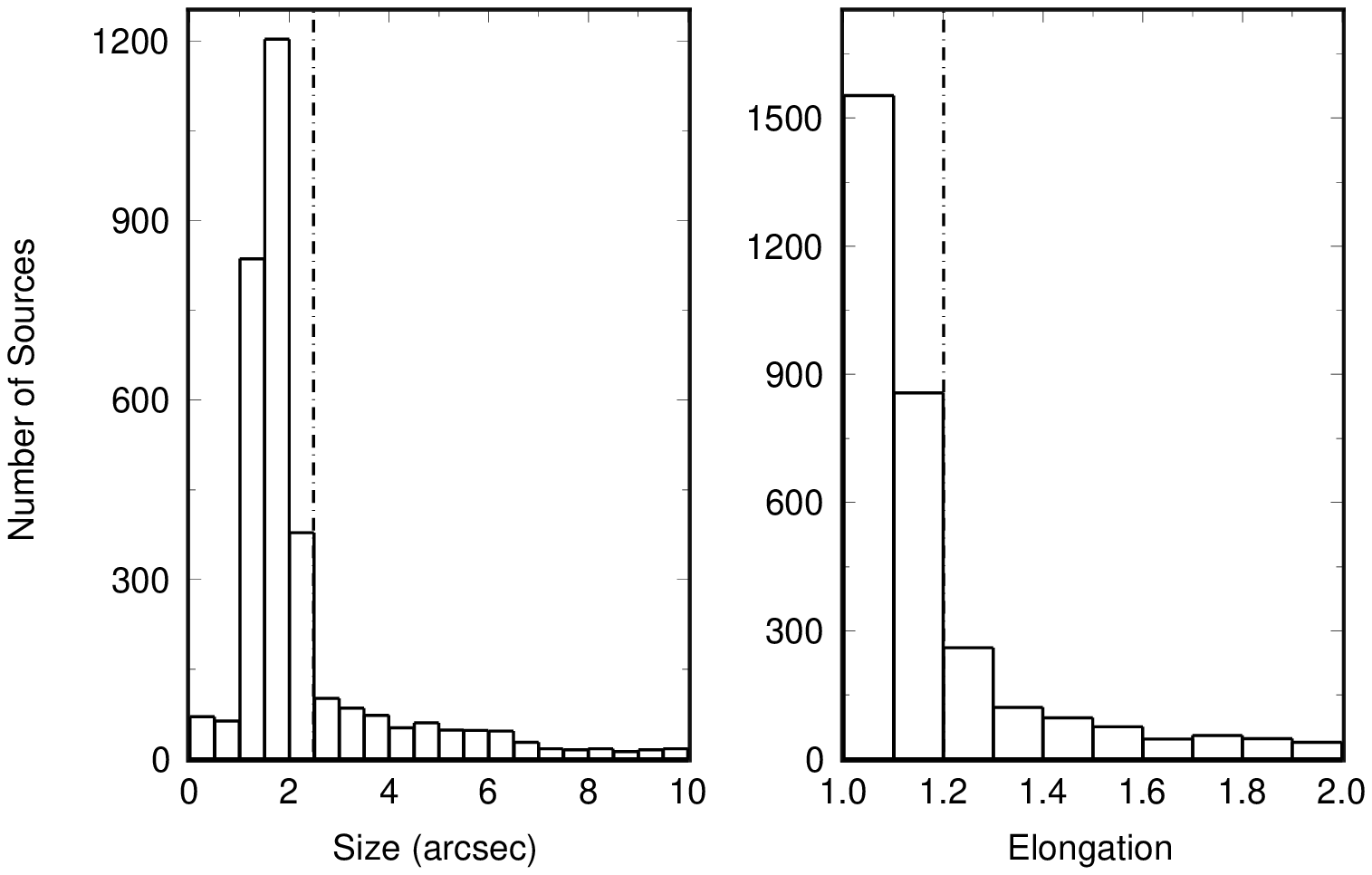}
\figcaption[SPark.fig3.eps]
{Size and elongation histograms for the fields of Taurus-Auriga class I 
sources.  Left panel: size histogram for 3176 sources detected on K-band
images of Taurus-Auriga protostars. Right panel: elongation histogram. 
The dot-dashed line in each panel indicates the division between point 
sources and extended sources.} 

\epsfxsize=7.0in
\hskip 10ex
\epsffile{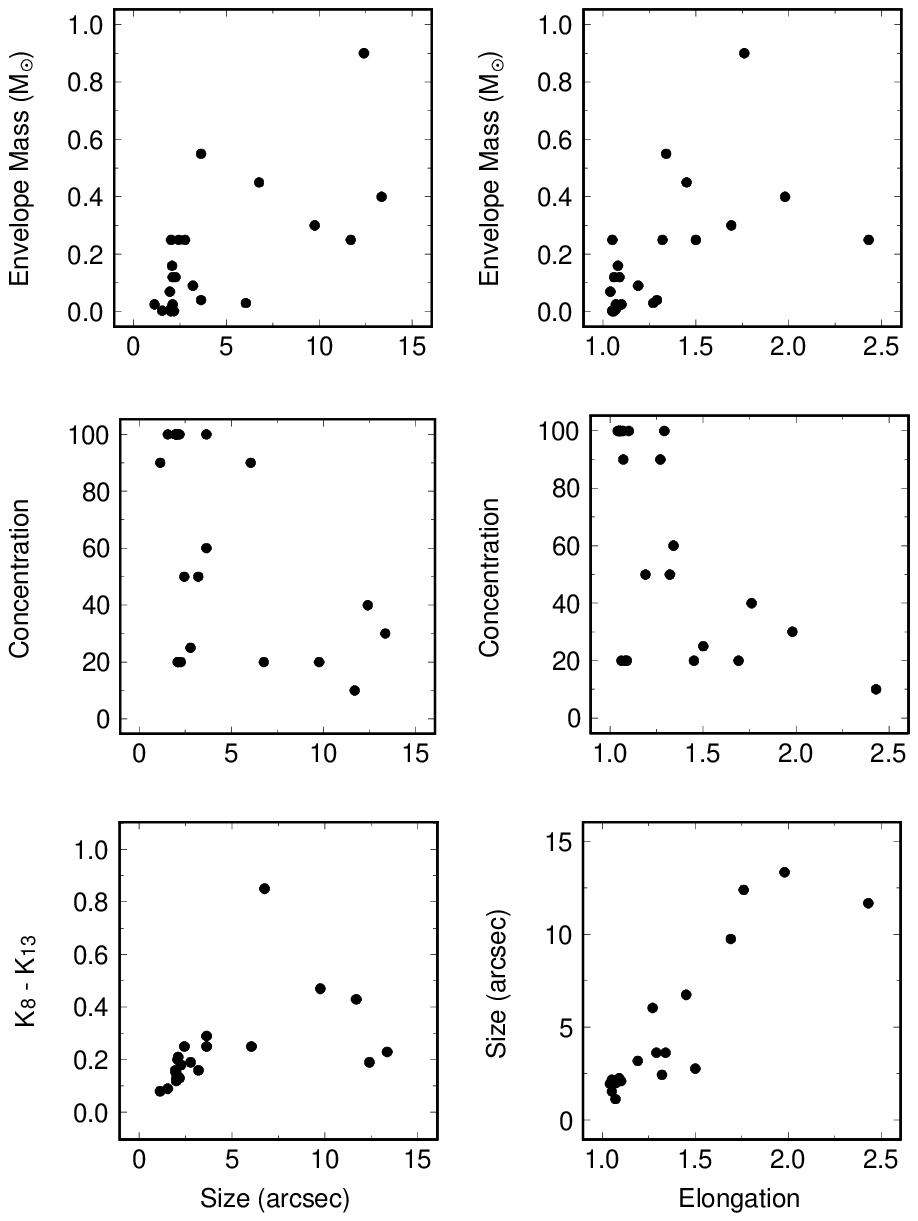}
\figcaption[SPark.fig4.eps]
{Correlations between K-band (size and elongation) and mm continuum properties
(envelope mass and concentration) of Taurus-Auriga class I sources. Source 
elongation correlates well with source size (lower right panel) and 
mm continuum concentration (middle right panel). The source size 
correlates well with mm continuum concentration (left middle panel).} 

\end{document}